\documentclass[aps,prb,twocolumn,superscriptaddress]{revtex4-1}
\usepackage{graphicx}
\usepackage{subfigure}
\usepackage{dcolumn}
\usepackage{bm}
\usepackage{hyperref}
\usepackage{indentfirst}
\usepackage{braket}
\usepackage{amsmath}
\hypersetup{colorlinks=true, citecolor=blue, urlcolor=blue, linkcolor=blue}
\pdfoutput=1

\begin{document}

\title{Enhancement of Chirality-Induced Spin Selectivity by Strong Electron Correlations}

\author{Meng Xu}
\affiliation{Department of Physics and State Key Laboratory of Surface Physics, Fudan University, Shanghai 200433, China}

\author{Yan Chen}
\email{yanchen99@fudan.edu.cn}
\affiliation{Department of Physics and State Key Laboratory of Surface Physics, Fudan University, Shanghai 200433, China}
\affiliation{Shanghai Branch, Hefei National Laboratory, Shanghai 201315, China}

\date{\today}

\begin{abstract}
Chirality-induced spin selectivity is a spin-splitting phenomenon from a helical structure with a considerably effective spin-orbit coupling. This unexpectedly large spin-splitting phenomenon has been experimentally observed in chiral organic molecules, which typically show a weak spin-orbit coupling. To understand this, we use the renormalized mean-field theory and Landauer-B\"{u}ttiker formulas to study the transport properties of single-stranded DNA in the presence of strong electron correlation. It shows a significant spin polarization of 46.5\% near the Coulomb repulsion limit, which explains the extremely high spin polarization observed in experiments. Compared to systems without electron correlation, the averaged spin polarization in this case is 2 to 4 times greater across various system sizes. Furthermore, the parameter dependence of the spin polarization and the underlying Metal-Insulator transition are studied.
\end{abstract}

\maketitle


\section{Introduction}
Early in some photoelectron transmission spectroscopy studies\cite{ray1999asymmetric, gohler2011spin}, people
discovered that electron transmission through a single layer of chiral molecules shows a preference for one specific spin orientation over the other at room temperature. This phenomenon, known as the Chirality-induced Spin Selectivity (CISS) effect, has opened up new possibilities for injecting spin-polarized current and fabricating spintronics devices without the need for a permanent magnetic layer\cite{dor2013chiral, qian2022chiral, kulkarni2020highly}. The CISS effect has also attracted intense interest in biochemistry for its application in enantiomer separation\cite{banerjee2018separation}. Apart from this, a recent study has proposed that the chiral symmetry breaking caused by the CISS effect could lead to enantioselective synthesis on the prebiotic Earth, leading to the homochiral assembly of life's essential molecules\cite{ozturk2022origins}.
To investigate the mechanisms of the CISS effect, numerous transport experiments on spin selectivity have been carried out\cite{xie2011spin,naaman2012chiral,naaman2015spintronics, li2020chiral}. The electrical behavior of organic self-assembled monolayers was examined in a two-terminal system. Spin polarization (SP) is calculated as the difference between the intensities of signals of opposite spin\cite{naaman2012chiral} $[I_{+} - I_{-}]/[I_{+} + I_{-}]$, where $I_{+}$ and $I_{-}$ are the intensities of the signals corresponding to the spin aligned parallel and antiparallel to the electrons' velocity, respectively. Among these studies\cite{naaman2015spintronics, li2020chiral}, it is believed that SP is caused by a strong effective spin-orbit coupling (SOC). However, spin splitting of organic molecules is usually negligible since the SOC of carbon is as weak as 3.4 meV\cite{steele2013large}. To understand this puzzling phenomenon, some single electron model studies\cite{sun2005quantum, gutierrez2012spin, matityahu2016spin} have proposed that a combination of SOC, helical symmetry, and nonequilibrium could lead to significant SP. On the other hand, it has been suggested that many-body and correlation effects can also improve the SP effect\cite{du2020vibration, zhang2020chiral, fransson2021charge}. However, current transport studies on many-body systems often use equations of motion based on perturbation theory. This approach may need to be more practical in cases of strong correlation. As a result, the impact of correlation effects on the CISS effect still needs to be determined.

In the present paper, we investigate the strong electron correlation effects of a B-type single-stranded DNA (ssDNA) molecule, described by an extended Hubbard model. By the method of renormalized mean-field theory\cite{zhang1988renormalised} (RMFT) and Landauer-B\"{u}ttiker formulas\cite{datta1997electronic}, we explore the impact of strong correlation on electronic transport and SP phenomena at electron half-filling. RMFT provides an intuitive understanding of the Hubbard model as it approaches the Coulomb repulsion limit, revealing significant SP as the system transitions to an insulator state. The kinetic effect is considerably suppressed at the strong correlation limit, and the SOC effect becomes prominent. Compared to systems without electron correlation, the averaged spin polarization in such case is 2 to 4 times greater across various system sizes, even in systems with weak SOC.

\section{Methodology and Models}
The possess of measuring charge transport through the right-handed ssDNA with an intramolecular electric field can be simulated by the Hamiltonian as $H=H_{\mathrm{mol}}+H_{\mathrm{lead}}+H_{\mathrm{c}}$, in which
\begin{equation} \label{eq1}
H_{\mathrm{mol}}=\sum_{i;\sigma}^{N-1}-t c_{i+1, \sigma}^{\dagger} c_{i, \sigma}+h.c.+U \sum_{i; \sigma}^{N} n_{i, \sigma} n_{i \bar{\sigma}} +H_{\text{so}}.
\end{equation}
Here, $H_{\mathrm{mol}}$ represents the one-dimensional Hubbard model, incorporating the spin-orbit coupling effect $H_{\text{so}}$. This model consists of $N$ building blocks of DNA, with $\sigma$ denoting the spin index $\{\uparrow, \downarrow\}$. The operator $c_{i, \sigma}^{\dagger}$ represents the creation of an electron with spin $\sigma$ at the $i$th site of the DNA. The hopping integral between nearest-neighbor sites is denoted by $t$, while $U$ represents the repulsive strength of electrons with different spins occupying the same site. Then, $H_{\mathrm{lead}}+H_{\mathrm{c}}=\sum_{i;\sigma} -t_{\mathrm{leads}} a_{i+1,\sigma}^{\dagger} a_{i,\sigma} - t_{\mathrm{junc}} a_{0,\sigma}^{\dagger} c_{1, \sigma} - t_{\mathrm{junc}} a_{N+1,\sigma}^{\dagger}c_{N,\sigma}+ h.c.$ describes the left $(i \textless 1)$ and right $(i \textgreater N)$ semi-infinite nonmagnetic electrodes, as well as the coupling $t_{\mathrm{junc}}$ between the electrodes and the DNA.

\begin{figure}[!htbp]
\centerline{\includegraphics[width=0.45\textwidth]{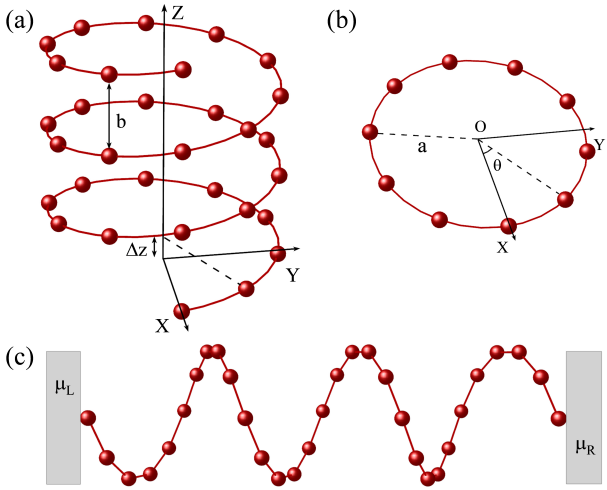}}
\caption{(a) Schematic view of a right-handed ssDNA molecule along the z direction. Here, $b$ denotes the pitch of spiral-arranged nucleobases. And $\Delta z$ is the stacking distance between neighboring base pairs. (b) Projection to the polar coordinate plane with molecule radius $a$ and twist angle $\theta$. (c) The schematic illustration of a chiral molecule connected to two nonmagnetic electrodes. Electrons are transported from the left electrode through the conductor to the right electrode. As the ssDNA molecule is a conductor, we set $a = 1$ nm, $b = 0.34$ nm, and $\Delta \theta = \pi/5$, typical values of B-form DNA.
\label{fig1} 
}
\end{figure}

For electrons moving with momentum $\mathbf{p}$ through the helix, the field induces a magnetic field in the electron’s rest frame, leading to SOC of electrons:
$H_{\text {so}}=\lambda \bm{\sigma}(\mathbf{p} \times \mathbf{E}_{\text {helix}})$. From Dirac’s equation\cite{BERESTETSKII1982118}, the SOC coefficient is $\lambda=e \hbar/4m_{0}^2 c^2$, and the Pauli matrices are $\bm{\sigma}=(\sigma_x, \sigma_y, \sigma_z)$. We consider electrons moving in a helical electrostatic field $\mathbf{E}_{\text {helix}}$\cite{hochberg1997representing, PhysRevE.52.901}. The electric field comprises electrically charged building blocks along the helical backbone of single-strand B-form DNA, shown in Fig.~\ref{fig1}(a). The following equation represents this electric field:
\begin{equation} \label{eq3}
    \mathbf{E}_{\text {helix}}=-E_{0} \sum_{l, m} g_{l, m}(z)[\cos ( m \theta ), \sin ( m \theta)].
\end{equation}
In this equation, the index $m \in [0, M)$ marks the $z$ coordinate of $M$ building blocks running along one helical turn. The index $l \in [0, L]$ marks the number of $L$ helical turns of the whole helix, separated by pitch $b$ in their $z$ coordinate. As shown in Fig.~\ref{fig1}(b), $\theta = 2\pi \Delta z /b$ is the twist angle between adjacent building blocks with stacking distance $\Delta z$. $E_{0}$ is a constant that is proportional to the local charge density, and $g_{l, m}(z)=a \left\{a^2+[(z-l b-m \Delta z)]^{2}\right\}^{-3 / 2}$ denotes the distance between the incident electron and each local charged building blocks.

In reality, it is a complex three-dimensional problem. For the sake of brevity, we assume $\mathbf{p}=(0,0,p_z)$, indicating that only the component responsible for transporting the electron through the DNA from one end to the other is being considered\cite{hochberg1997representing,gutierrez2012spin}. When the electron moves through the backbone of the molecule, it senses a $z$-coordinate-dependent electric field $\mathbf{E}_{\text {helix}}$. Thus the SOC is reduced to $H_{\mathrm{so}}=\lambda\left(-\sigma_{x} p_{z} E_{y}+\sigma_{y} p_{z} E_{x}\right)$. After the discretization treatment\cite{SM} of Hamiltonian for $H_{\mathrm{so}}$, we got the matrix representation:
\begin{equation} \label{eq4}
\begin{aligned}
    H_{\mathrm{so}} &=\sum_{i}^{N} \frac{\alpha}{2 \Delta z} {\left[ d(z_{i}) c_{i+1, \uparrow}^{\dagger} c_{i, \downarrow} - d^{*}(z_{i}) c_{i+1, \downarrow}^{\dagger} c_{i, \uparrow} \right.} \\ 
     &{\left. - d(z_{i}) c_{i, \uparrow}^{\dagger} c_{i+1, \downarrow} + d^{*}(z_{i}) c_{i, \downarrow}^{\dagger} c_{i+1, \uparrow} \right]}\\
    &+\sum_{i}^{N}  \alpha \left[f(z_{i})c_{i, \uparrow}^{\dagger}  c_{i, \downarrow}-f^{*}(z_{i}) c_{i, \downarrow}^{\dagger}  c_{i, \uparrow}\right]
\end{aligned},
\end{equation}
where $\alpha=\hbar \lambda E_0$, $d(z_i)=\sum_{l, m} e^{\mathrm{-i} m \theta} g_{l, m}(z_i)=\sum_{l, m} e^{\mathrm{-i} m \theta} a \left\{a^2+[(z_i-l b-m \Delta z)]^{2}\right\}^{-3 / 2}$ and $f(z_i)=\partial_{z_i} d(z_i)$ denote the SOC term. It is important to mention that the SOC, in this case, is derived from the helical structure rather than just the individual atoms. The SOC consists of two parts: one is the spin-flip between nearest neighbor lattices, and the other is the on-site spin flip. The latter introduces a non-Hermitian nature to the system, allowing the system to exhibit SP\cite{matityahu2016spin}. Since Eq.~\eqref{eq4} exhibits time-reversal symmetry, the total conductance will remain unpolarized due to the phase-locking effect. As previously proposed by Sun et al.\cite{PhysRevB.71.155321}, one cannot solely rely on SOC to generate SP in a two-terminal setup.

As in Fig.~\ref{fig1}(c), the molecule is assumed to be attached to two semi-infinite electrodes depicted by a tight-binding Hamiltonian with the hopping integral $t_{\mathrm{junc}}$. Landauer-B\"{u}ttiker formulas\cite{datta1997electronic} then give a linearized relationship between the currents flowing through the electrodes for small biases (assume temperature is zero): $I_{p \alpha}=\frac{e^{2}}{h} \sum_{q \alpha '} T_{p \alpha, q \alpha^{\prime}}\left(E_{F}\right)\left(V_{p}-V_{q}\right)$, where $T_{p \alpha, q \alpha^{\prime}}$ is the transmission coefficient for electrons from electrode $q$ with spin $\alpha '$ to electrode $p$ with spin $\alpha$. As such, the current-voltage characteristics of the device can be entirely determined by calculating the transmission coefficients between all electrodes. We formulate the Landauer approach using Green's function formalism\cite{ryndyk2016theory} $T_{p \alpha , q \alpha^{\prime}} = \operatorname{Tr}\left[\boldsymbol{\Gamma}_{p \alpha} \mathbf{G}^{r} \boldsymbol{\Gamma}_{q \alpha^{\prime}} \mathbf{G}^{a}\right]$. The retarded and advanced Green's function is $\mathbf{G}^{r}=\left[\mathbf{G}^{a}\right]^{\dagger}=\left[E \mathbf{I}-\mathbf{H}_{\mathrm{mol}}-\mathbf{H}_{\mathrm{so}}-\sum_{p \alpha} \boldsymbol{\Sigma}_{p \alpha}^{r} \right]^{-1}$. $\boldsymbol{\Sigma}_{p \alpha}^{r}$ is the retarded self-energy of $p$th electrode with spin $\alpha$ due to the coupling to the molecule. And the level-width function $\boldsymbol{\Gamma}_{p \alpha}= i \left[\boldsymbol{\Sigma}_{p \alpha}^{r}-\boldsymbol{\Sigma}_{p \alpha}^{a}\right]$ represents the coupling of electrons between the DNA and the electrodes. All effects of the equilibrium electrodes can be included in the molecule matrix through self-energy. The total transmission function for the system can be written as
\begin{align} \label{eq9}
    T(E)&= \Gamma_{\uparrow}^{R}\left(\Gamma_{\uparrow}^{L}\left|G_{1 \uparrow, N \uparrow}\right|^{2}+\Gamma_{\downarrow}^{L}\left|G_{1 \downarrow, N \uparrow}\right|^{2}\right) \nonumber \\ &+\Gamma_{\downarrow}^{R}\left(\Gamma_{\uparrow}^{L}\left|G_{1 \uparrow, N \downarrow}\right|^{2}+\Gamma_{\downarrow}^{L}\left|G_{1 \downarrow, N \downarrow}\right|^{2}\right) \nonumber \\
    &= T_{\mathrm{up}}(E)+T_{\mathrm{down}}(E). 
\end{align}

Note that we calculate the spin transmission spectra from the left electrode to the right one. The spin-up and spin-down transmission coefficients at the right electrode are $T_{\mathrm{up}}$ and $T_{\mathrm{down}}$. The SP in the ssDNA molecule is defined as 
\begin{equation} \label{eq10}
    P_s(E)=\frac{T_{\mathrm{up}} -T_{\mathrm {down }} }{T}.
\end{equation}
We focus on examining the influence of strong electronic correlation effects on electron transport and SP phenomena.

To investigate the strong correlation effects, we study Laughlin's idea of the variational wave function method, which uses the BCS wave function to describe the Gossamer superconductor state\cite{laughlin2002gossamer}. In this work, we use the RMFT method\cite{zhang1988renormalised}. Although it is somewhat different, the trial wave function is selected as the noninteracting electron wave function with the spin component, $|\Psi_0 \rangle=\sum_{n \sigma} \psi_{n \sigma} c_{n \sigma}^{\dagger}|0\rangle$. The state of this correlated system is a generalization of the partially projected noninteracting electron state.

\begin{equation} \label{eq5}
\begin{gathered}
    \left|\Psi\right\rangle=\Pi_{\nu}\left|\Psi_{\mathrm{0}}\right\rangle, \\
    \Pi_{\nu}=\prod_{i}\left(1-\nu n_{i \uparrow} n_{i \downarrow}\right)
\end{gathered}.
\end{equation}
$\Pi_{\nu}$ is a projection operator with the parameter $\nu$ determined by optimizing the energy by a variational calculation. When $\nu=0$, there is an uncorrelated state. And if $U \rightarrow \mathbf{\infty}$, which means $\nu=1$, corresponds to the limit of no doubly occupied state. The parameter $\nu$ ranges from 0 to 1 and changes according to the value of $U$, which accounts for electron correlation by partially projecting out doubly occupied sites. The projection operator is considered by renormalized factors, which were proposed by Gutzwiller\cite{gutzwiller1965correlation, vollhardt1984normal}. The renormalized factors $g_t$, $g_{\mathrm{so_1}}$, and $g_{\mathrm{so_2}}$ are determined by the ratio of the probabilities of the physical processes in the states $|\Psi\rangle$ and $|\Psi_0 \rangle$. Following the counting arguments\cite{edegger2007gutzwiller} of classical statistical weighting factors, we have
\begin{equation} \label{eq7}
\begin{gathered}
        g_{\mathrm{s o_1}} = g_{\mathrm{t}} = \frac{(n-2 d)(\sqrt{d}+\sqrt{1-n+d})^{2}}{(1-n / 2) n}, \\ 
        g_{\mathrm{s o_2}} = \frac{n-2d}{n\left(1-\frac{n}{2}\right)},
\end{gathered}
\end{equation}
where $d=\langle n_{i \uparrow} n_{i {\downarrow}} \rangle$ is the average electron double occupation number. With $n= n_{i \uparrow} + n_{i {\downarrow}}$, the electron density $n \leq 1$, and $0 \leq d \leq 1/4$. The expectation value of hopping and the SOC in the state $|\Psi \rangle$ are related to those in the partially projected state $|\Psi_0 \rangle$
\begin{subequations}
\begin{align} 
    \langle c_{i+1,\sigma}^{\dagger} c_{i,\sigma} \rangle &= g_{\mathrm{t}} \langle c_{i+1,\sigma}^{\dagger} c_{i,\sigma}\rangle_0 \label{eq6a}, \\
    \langle c_{i+1,\sigma}^{\dagger} c_{i, \bar \sigma} \rangle &= g_{\mathrm{so_1}} \langle c_{i+1,\sigma}^{\dagger} c_{i,\bar \sigma}\rangle_0 \label{eq6b}, \\ 
    \langle c_{i,\sigma}^{\dagger} c_{i,\bar \sigma} \rangle &= g_{\mathrm{so_2}} \langle c_{i,\sigma}^{\dagger} c_{i,\bar \sigma}\rangle_0. \label{eq6c}
\end{align}
\end{subequations}
Eq.~\eqref{eq6b} represents the SOC energy coming from $d(z)$ terms in $H_{\mathrm{so}}$, which corresponds to the nearest neighbor spin-flip. Similarly, Eq.~\eqref{eq6c} refers to the onsite spin-flip SOC energy that comes from $f(z)$ terms in $H_{\mathrm{so}}$.

We now proceed with the variational calculation to determine the parameter $\nu$. The variational energy per site is given by $E = \langle \Psi|H|\Psi \rangle = U d + \langle H_\mathrm{t} \rangle + \langle H_{\mathrm{so}} \rangle$ on the correlated basis. Using the Gutzwiller approximation, we can define a renormalized Hamiltonian based on an uncorrelated state $|\Psi_0 \rangle$:
\begin{equation} \label{eq8}
    H_{\mathrm{mol}}^{'} = U d + g_{\mathrm{t}} H_{\mathrm{t}} + g_{\mathrm{s o_1}} H_{\mathrm{s o}} + g_{\mathrm{s o_2}} H_{\mathrm{s o}}.
\end{equation}
The renormalized Hamiltonian approach transforms the original variational parameter $\nu$ into the variational parameter $d$. There is a one-to-one correspondence between $\nu$ and $d$\cite{gutzwiller1965correlation,vollhardt1984normal}. And the variational energy becomes $E = U d + g_{\mathrm{t}} \langle H_{\mathrm{t}} \rangle_0 + g_{\mathrm{s o_1}} \langle H_{\mathrm{s o}} \rangle_0 + g_{\mathrm{s o_2}} \langle H_{\mathrm{s o}} \rangle_0$. Note that renormalized factors are functions of $d$. By optimizing the energy $\partial E(d) / \partial d = 0$, the value of the variational parameters $d$ and the renormalized factors are determined. If $U=0$, Eq.~\eqref{eq7} has $d=0.25$. So, all renormalization factors come to 1. It is an uncorrelated system that allows two electrons per site. If in the strong correlation limit $U \rightarrow \infty$, double occupation is forbidden. We have $g_{\mathrm{t}} \rightarrow 0$ and $g_{\mathrm{s o_2}} \rightarrow 2$. At that time, even though SOC is weak in biological molecules, it dominates over kinetic terms.

We have considered the realistic parameters of the B-form ssDNA molecule\cite{wing1980crystal}, the radium $a = 1$ nm, pitch $b = 0.34$ nm, twist angle $\Delta \theta = \pi/5$ and the number of base pairs (bp) $N=30$. As described by James Watson and Francis Crick, the B-form DNA is believed to be the predominant form in cells\cite{richmond2003structure}. For B-DNA with a C-G base pair, DFT calculations\cite{kubar2008efficient} have shown that $t$ ranges from 30 to 60 meV. In this study, we adopt a value of $t=30$ meV. When comparing the order of SOC to $t$, it is evident that the SOC is significantly smaller\cite{steele2013large}. It has been roughly estimated that $\alpha$ lies between 1 and 6 meV nm\cite{gutierrez2012spin, sun2005quantum}. The matrix elements of the level-width functions for the attached nonmagnetic electrodes, $\Gamma_L$ and $\Gamma_R$, are equivalent to $\Gamma$. Typically, their values are comparable to that of $t$. Various scenarios have also been examined, extending from the weak coupling limit to the strong coupling limit\cite{mehrez2005interbase} between the DNA and the electrodes.

\section{Numerical Results}

\begin{figure}[!htbp]
\centerline{\includegraphics[width=0.5\textwidth]{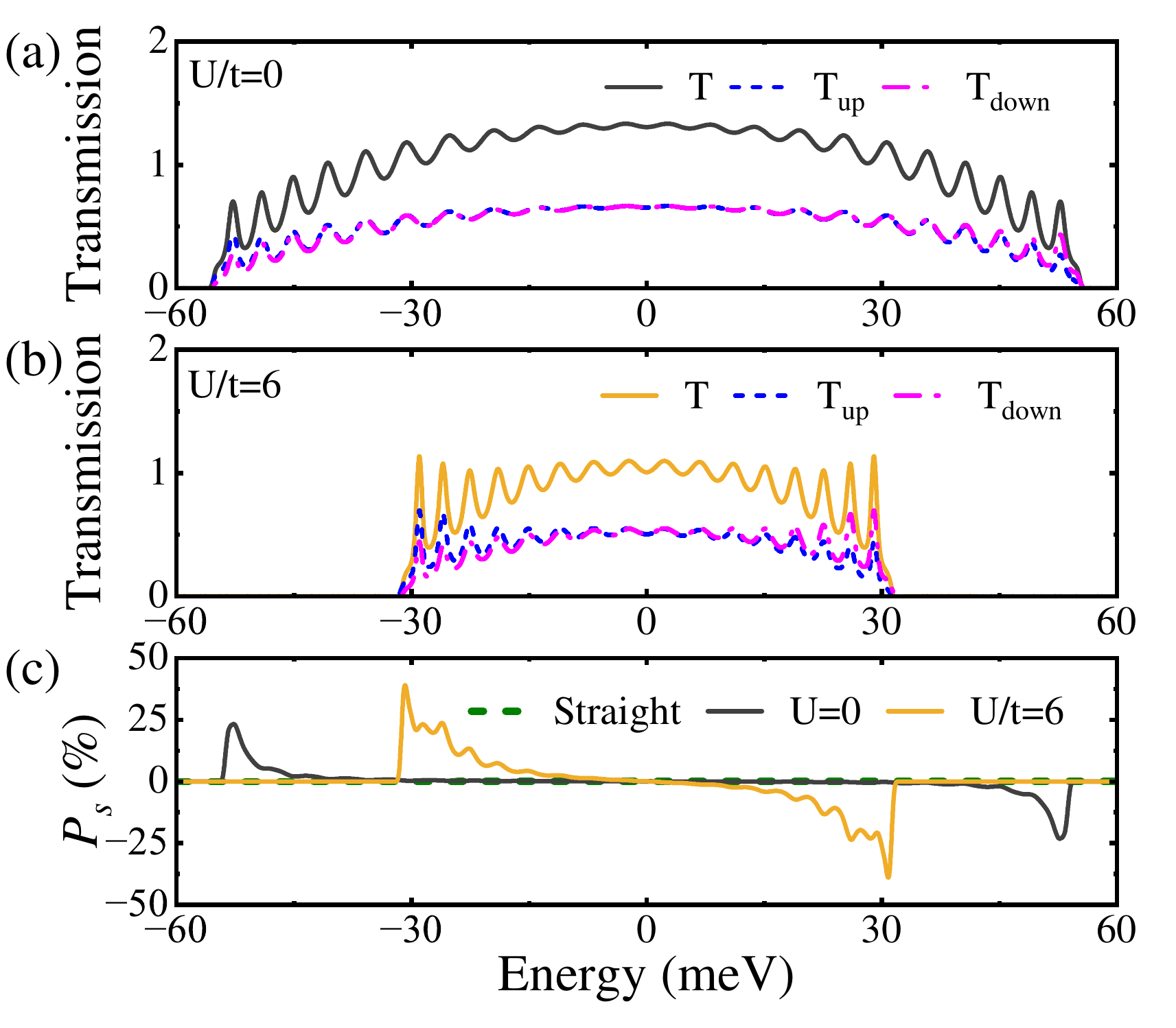}}
\caption{The transmission function $T$ and the SP $P_s$ of electrons exhibit energy-dependence. In (a) and (b), including the transmission functions $T_\mathrm{up}$ and $T_\mathrm{down}$ for different spins, as well as their sum $T$, plotted as a function of Fermi energy. Specifically, (a) corresponds to $U/t=0$ while (b) corresponds to $U/t=6$. (c) The behavior of $P_s$ for electrons with various $U$. The solid line demonstrates that coulomb repulsion significantly enhances the electron's SP for the right-handed B-type DNA mentioned in the main text, characterized by a twist angle of $\Delta \theta=\pi/5$ and a pitch of $b=0.34$ nm, as referenced in \cite{wing1980crystal}. The dashed line represents a zero SP for the straight chain regardless of $U$ values, with base pairs separated by a distance of 0.34 nm. All the cases here have $N=30$, $t=30$ meV, $\alpha=1$ meV nm and $\Gamma=50$ meV. 
\label{fig2} 
}
\end{figure}
           
Modeling the external electrodes allows us to scan the system's transmission profile over a spectrum. When ballistic transport occurs with spin degeneracy, the envelope of the transmission function $T$ equals 2. However, due to the finite system with open boundary conditions, the calculated value of $T$ is slightly below 2, as shown in Fig.~\ref{fig2}~(a). Additionally, contact resistance prevents ballistic transport from the system when there is an energy mismatch between the electrodes and the central scattering region. Furthermore, electron scattering occurs when $U$ is involved, which usually electrodes to a decrease in $T$\cite{PhysRevB.52.R17040}, except for some special cases, such as the single-impurity Anderson model at the particle-hole symmetry point\cite{PhysRevB.105.165120}.

The effect of Coulomb repulsion on transmission can be seen in Fig.~\ref{fig2}~(a), (b). When $U/t$ increases from 0 to 6 in a weak SOC, the energy window gradually shrinks, and the transmission amplitude decreases. Fig.~\ref{fig2}~(b) shows the transmission functions for several spin channels, with $\alpha=1$meV nm and $U/t=6$. The solid line represents the total transmission $T$ for all spin channels. The dashed-dotted line represents $T_\mathrm{down}$, which means the spin $\mathbf{S}$ of the electron is antiparallel to the momentum $\mathbf{p}$ from left to right electrodes. Similarly, the dashed line $T_\mathrm{up}$ represents electrons propagating with $\mathbf{S}$ and $\mathbf{p}$ in the same direction. For the electron-like band $(E \textless 0)$, it was found that there are more electrons with spin-up than spin-down moving from the left electrode to the right electrode, resulting in a positive $P_s$. Fig.~\ref{fig2}~(c) shows that the SP effect is significant for energies nearing the band edge. When Coulomb repulsion $U/t=6$, the largest $P_s$ is almost twice the value of the uncorrelated system. The correlation effects result in a shrinkage and renormalization of the energy band, which could significantly enhance the spin filtering effect. It helps to understand the experimentally observed significant spin splitting in molecular systems of CISS\cite{gohler2011spin}. Besides, there is no SP for a straight chain regardless of the values of $\alpha$ and $U$, as indicated by the green dashed line in Fig.~\ref{fig2}~(c). This highlights the importance of a helical structure for SP.

\begin{figure}[t]
\centerline{\includegraphics[width=0.5\textwidth]{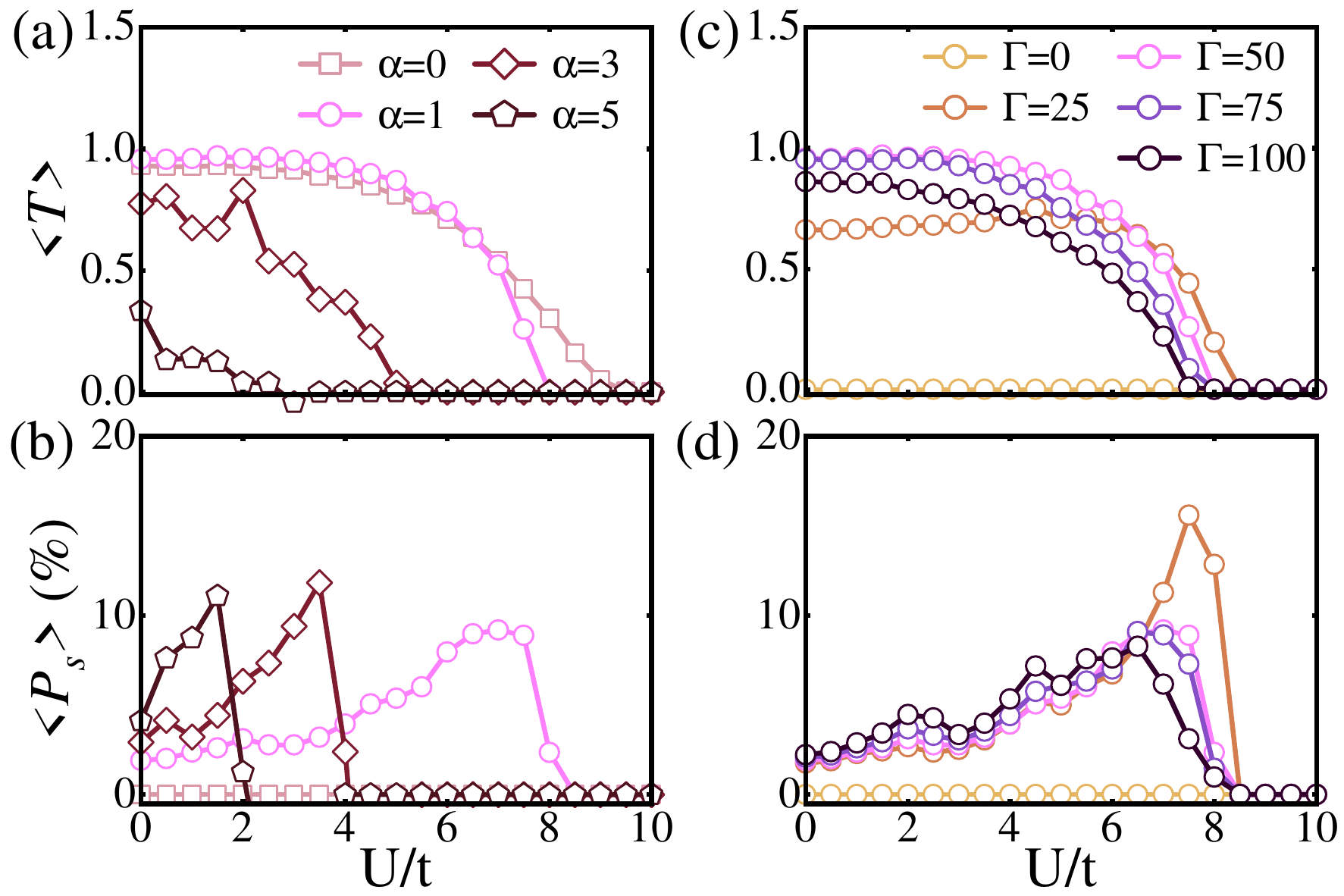}}
\caption{The averaged transmission and spin polarization ($\langle T \rangle$ and $\langle P_s \rangle$) were calculated by averaging $T$ and $P_s$ over the electron-like band ($E<0$). In (a) and (b), these values were obtained for various strengths of spin-orbit coupling $\alpha$, with fixed parameters of $N=30$ and $\Gamma=50$ meV. Similarly, in (c) and (d), are $\langle T \rangle$ and $\langle P_s \rangle$ for different level-width functions $\Gamma$ of non-magnetic electrodes, with  $N=30$, $\alpha=1$meV nm. Here $t=30$ meV, the unit of $\alpha$ is meV nm, and the unit of $\Gamma$ is meV.
\label{fig3} 
}
\end{figure}
Since we focus on the electron-like band, the averaged $\langle P_s \rangle$ can be calculated as follows: $\langle P_s \rangle = [\langle T_{\mathrm{up}} \rangle - \langle T_{\mathrm {down}} \rangle ] / \langle T \rangle$, where $\langle T \rangle$ is the averaged transmission function over the electron-like band. In Fig.~\ref{fig3}~(a), we observe that $\langle T \rangle$ decreases with an increase in $\alpha$. This is due to the influence of parameters related to SOC, such as $\frac{\alpha d\left(z_i\right)}{2 \Delta z}$ and $\alpha f\left(z_i\right)$, which vary with the position $z$. The latter parameter introduces a non-uniform, non-Hermitian characteristic to the system, leading to a scattering in electron transport. Although the $\langle T \rangle$ is relatively small, as shown in Fig.~\ref{fig3}~(b), there is always a significant SP when the appropriate $U$ is chosen. The same applies to the effect of $U$ when looking at various $\Gamma$, representing the coupling between electrodes and the molecule. In Fig.~\ref{fig3}~(c), (d), SOC strength is fixed at 1 meV nm. When $\Gamma=0$, no signal can be detected at the electrodes because the system is disconnected. As $\Gamma$ rises to about 50 meV, $\langle T \rangle$ reaches its maximum. This means that the potential barrier at the contact interface is small. If the strength of $\Gamma$ increases further, the coupling will have a negative impact on the electrons transitioning into the molecule. In Fig.~\ref{fig3}~(d), it can be found that the $\Gamma$ of non-magnetic electrodes has no significant effect on SP.

\begin{figure}[!htbp]
\centerline{\includegraphics[width=0.5\textwidth]{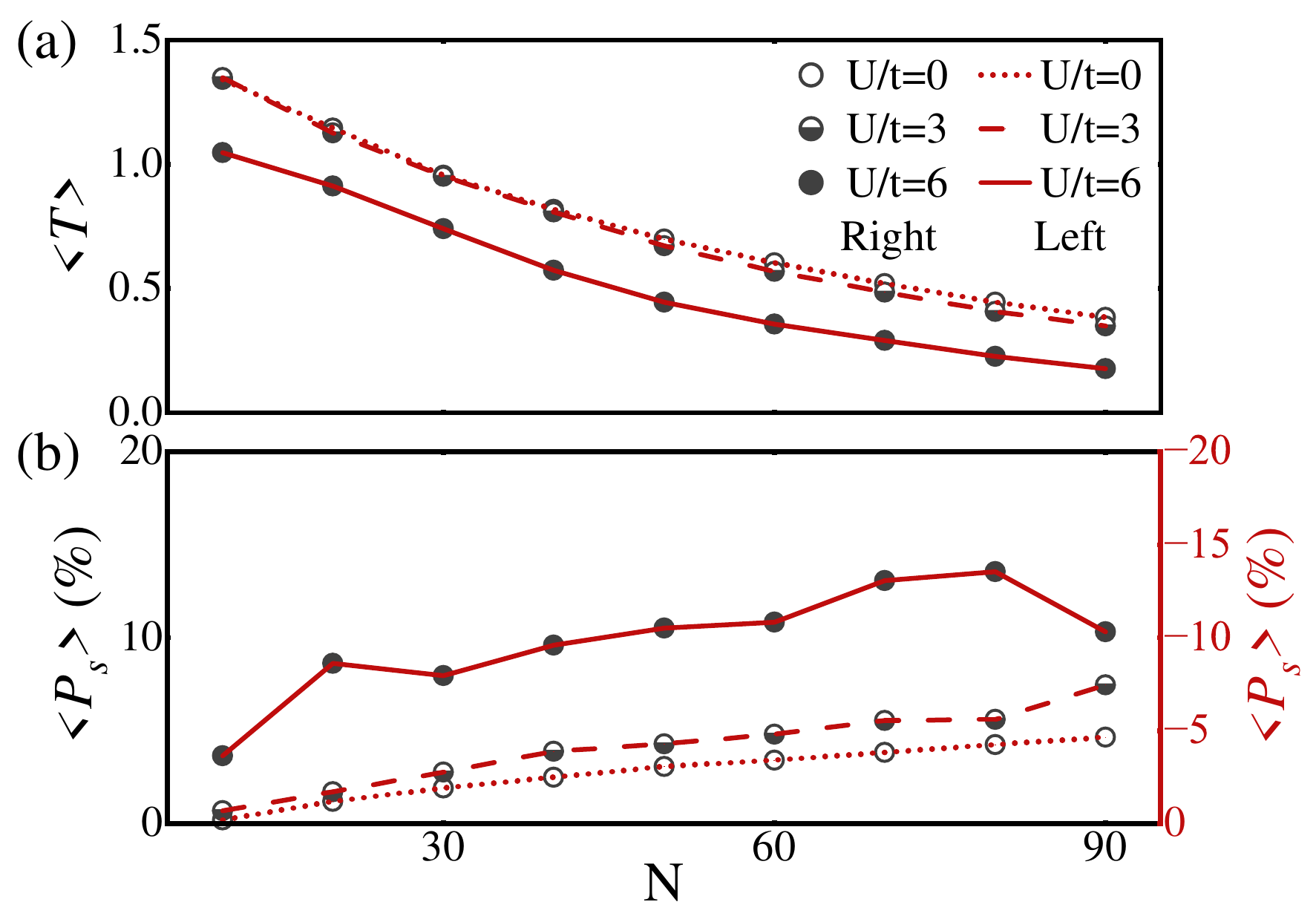}}
\caption{(a) $\langle T \rangle$ and (b) $\langle P_s \rangle$ with the opposite molecular chirality for different molecule length $N$, as well as their relation on $U$. The grey symbol represents a right-handed helix; the red line represents a left-handed helix. With $\Gamma=50$ meV and $\alpha=1$ meV nm.
\label{fig4} 
}
\end{figure}

The CISS effect also relates to the molecular structure itself. The green line in Fig.~\ref{fig1}~(c) has already demonstrated the critical role of the helical structure. In addition to this, the length and chirality of the molecule are also important, as shown in Fig.~\ref{fig4}~(a), (b). When the molecule length increases, $\langle T \rangle$ decreases slowly due to the increasing scattering length. But the $\langle P_s \rangle$ increases by $N$, which is consistent with the results of the experiment\cite{xie2011spin,gohler2011spin}. The grey open symbol in Fig.~\ref{fig4} represents the right-handed molecule, while the red line represents the left-handed molecule. We found that changing the molecular chirality electrodes to an opposite sign of the SP value, in accordance with the theory mentioned in a previous study\cite{PhysRevB.93.155436} and the experiment results\cite{naaman2015spintronics,clever2022}. And the transmission $T$ of molecules with different chirality remains the same regardless of $N$.

Furthermore, it is crucial to highlight the importance of $U$. As $U$ increases in Fig.~\ref{fig4}~(a), electron scattering increases. When $U$ approaches the critical value $U_c$, $\langle T \rangle$ for any given $N$ approaches zero, indicating the metal transition into a Mott insulator. As shown in Fig.~\ref{fig4}~(b), $\langle P_s \rangle$ increases with $U$ until it reaches a maximum and rapidly drops to zero upon the transition occurrence. Notably, for a special case of $\alpha=0$, the Metal-Insulator transition has been examined using Brinkman-Rice theory\cite{brinkman1970application}. According to this theory, the energy for a half-filled 1D chain is expressed as $E=U d-g_{t} \sum 2 t\left|\cos k\right|=U d-32t d(1-2 d)/\pi$. From this, we deduce that the critical value $U_c/t = 32/\pi$. Although the system with $N=30$ and $\alpha=0$ exhibits a lower $U_c$ than expected. After a finite size extrapolation of the critical point at various system sizes\cite{SM}, it is observed that $U_c$ increases and converges to a value of $32/\pi$, in agreement with the Brinkman-Rice theory. This indicates that electron correlation always results in a transition to an insulating state in a one-dimensional system, accompanied by an increase in the SP.

\section{Conclusion}
We propose an effective Hamiltonian using the RMFT method from an extended Hubbard model. We employ the Landauer-B\"{u}ttiker formulas to simulate electron transport through the electrode-ssDNA-electrode system. As $U$ indirectly affects SP by directly renormalizing SOC, this two-terminal system exhibits significant SP driven by $U$ near the transition to an insulator state. For a chiral organic molecule that typically exhibits weak SOC, the Coulomb repulsion with $U/t=6$ promotes the SP $P_s$ to reach 46.5\%. Compared to systems without electron correlation, the averaged $\langle P_s \rangle$ in this case is 2 to 4 times greater across various system sizes. This explains the extremely high SP observed in experiments\cite{gohler2011spin,xie2011spin}. Additionally, the SP grows with the length of the system, which is consistent with the results of the experiments\cite{naaman2015spintronics,clever2022}.

In contrast to variables studied before, we suggest a new approach to achieving CISS on a system with low SOC by introducing electron correlation. We demonstrate that the correlation strength can significantly control the level of spin polarization even without strong SOC. It is interesting to note that the concept of CISS provides a way to manipulate the spin current without needing a magnetic field, opening up possibilities for using chiral materials in spintronic devices. Furthermore, problems like the opposite sign of SP in molecules\cite{clever2022} with different attached groups (either at the C-end or N-end) on the interface can be studied based on our result.

\section{Acknowlegements}
We are grateful to Zhiyu Dong for many useful discussions. This work is supported by the National Key Research and Development Program of China Grant No. 2022YFA1404204, and the National Natural Science Foundation of China Grant No. 12274086.

\nocite{*}
\bibliography{RMFT}

\end{document}